# Reinforcement of Climate Hiatus by Decadal Modulation of Daily Cloud Cycle

Jun Yin[1,2], Amilcare Porporato[1,2*]

[1]Department of Civil and Environmental Engineering, Princeton University, Princeton, New Jersey, USA.

[2]Princeton Environmental Institute, Princeton University, Princeton, New Jersey, USA.
*Correspondence to: aporpora@princeton.edu

**Based on observations and climate model results, it has been suggested that the recent slowdown of global warming trends (climate hiatus), which took place in the early 2000s, might be due to enhanced ocean heat uptake[1–8]. Here we suggest an alternative hypothesis which, at least in part, would relate such slowdown to unaccounted energy reflected or reemitted by clouds. We show that the daily cloud cycle is strongly linked to pacific decadal oscillation (PDO) and that its decadal variations during the climate hiatus have an overall cooling effect. Such an effect may have partially, and temporarily, counteracted the greenhouse warming trends.**

The slowdown of global warming in the early 21th century[9], referred to as the climate hiatus, raises growing political and public concerns[10]. Observations and climate modelling results suggest that such a phenomenon is caused by the compounding effects of inter-annual and decadal variations of ocean circulation, aerosols, volcanic eruptions, and variation of solar irradiance[11–14]. While enhanced ocean heat uptake is regarded as one of its primary causes of the recent climate hiatus[1–4], there are still debates over which parts and depths of the ocean may be responsible for absorbing the imbalanced energy[5–8]. Such uncertainties stem in part from the temporal interpolation method used for satellite calibration and the sparse spatial/temporal sampling of the ocean heat content measurement[15–18]. It is thus logical to wonder whether the estimation of Earth's energy balance might have missed some energy component linked to the finer temporal resolutions (e.g. sub-daily timescale).

While the importance of seasonal cycles of clouds and radiative fluxes is widely acknowledged, the impact of daily cycles on climate could be even stronger (see Supplementary Fig. 1)[19–23]. Thus, variations in the daily cycle of clouds (DCC) have the potential to affect the Earth's energy balance and contribute to the climate variability. To test this hypothesis, we began by investigating the pacific decadal oscillation (PDO), which is widely regarded as an indicator of the climate hiatus[2,11,12,24] and is strongly linked to the Earth's mean surface temperature (see Supplementary Fig. 2). The PDO-temperature trends roughly show three periods of variations during the early 21th century. During the pre-hiatus (2000-2003) and post-hiatus (2013-present) periods, both Earth surface temperature and PDO increase; during the mid-hiatus (2003-2013) period, the PDO decreases while temperature keeps relatively constant. To explore whether DCC is also linked to the PDO, we used satellite observations of Single Scanner Footprint (SSF) from Clouds and the Earth's Radiant Energy System (CERES)[25] (see Methods), which can be used for long-term cloud trends analysis[18]. We focus on the cloud fraction, one of the most important cloud properties that are critical to the Earth's energy balance[9]. In general, the daytime clouds tend to reflect more solar radiation and cool the Earth, while the nighttime clouds tend to keep longwave radiation and warm the Earth. The daytime cloud fraction ($f_d$) was found to decrease during the pre- and post-hiatus periods, possibly contributing to the fast increase of PDO and global warming (see Fig. 1a); on the contrary, nighttime cloud fraction ($f_n$) was found to decrease during the mid-hiatus period, lowering the PDO index and partially cancelling the greenhouse gas effects (see Fig. 1b). A detailed daily cycle of cloud fraction at the hourly timescale, shown in Fig. 2 a, d, and g, reveals a similar pattern of alternative change of clouds during the transition of these three periods.

The link between DCC and PDO becomes even more evident when comparing the daily cloud amplitude ($f_a = f_n/2 - f_d/2$) with the daily mean cloud fraction ($f_0 = f_n/2 + f_d/2$) (see Methods). The amplitude $f_a$ combines the effects of daytime and nighttime clouds, while in the daily mean $f_0$ such opposite effects cancel out. As a result, $f_a$ becomes strongly correlated with PDO (Fig. 1c), while $f_0$ tends to be uncorrelated with PDO (Fig. 1d). The existence of such a strong correlation points toward potential contributions of DCC to the climate variability.

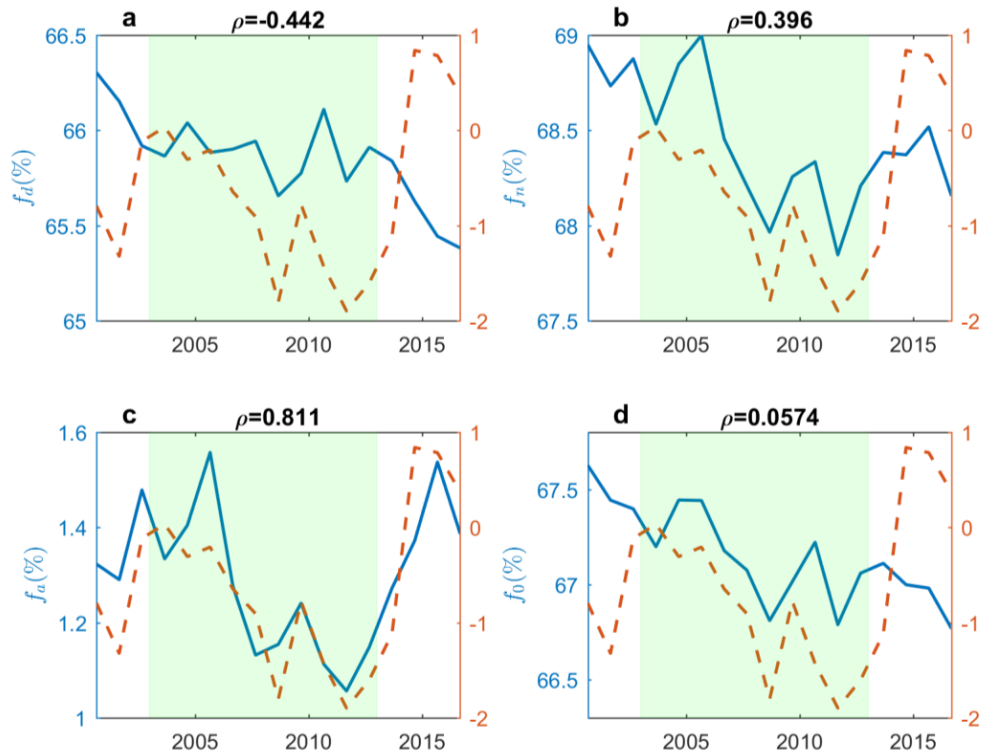

Fig. 1. Comparison of global mean cloud fraction with pacific decadal oscillation. The solid lines are the time series of (**a**) daytime, (**b**) nighttime, (**c**) daily amplitude, and (**d**) daily mean global cloud fraction; the dash lines are the pacific decadal oscillation (PDO) index. The correlation coefficients ($\rho$) of the two displayed data sets is reported on top of each panel. The shaded area divides the early 21st century into pre/mid/post-hiatus periods. Cloud fraction data are from SSF

CERES and PDO index are from NOAA National Centers for Environmental Information (see Methods).

A detailed spatial analysis in Fig. 2 shows that DCC also tends to shift the geographical patterns of global energy budget at different periods of climate hiatus. In particular, the long-term trends of daytime clouds and shortwave radiative fluxes at the top of the atmosphere show similar spatial patterns, especially over the Pacific Ocean (see Fig. 2 b, c, e, f, h, and i). Regions with positive trends of clouds and radiative fluxes in the pre- and post-hiatus periods likely shift into negative trends during the mid-hiatus period. Similar spatial patterns can be observed between nighttime clouds and longwave radiative fluxes (see Supplementary Fig. 3).

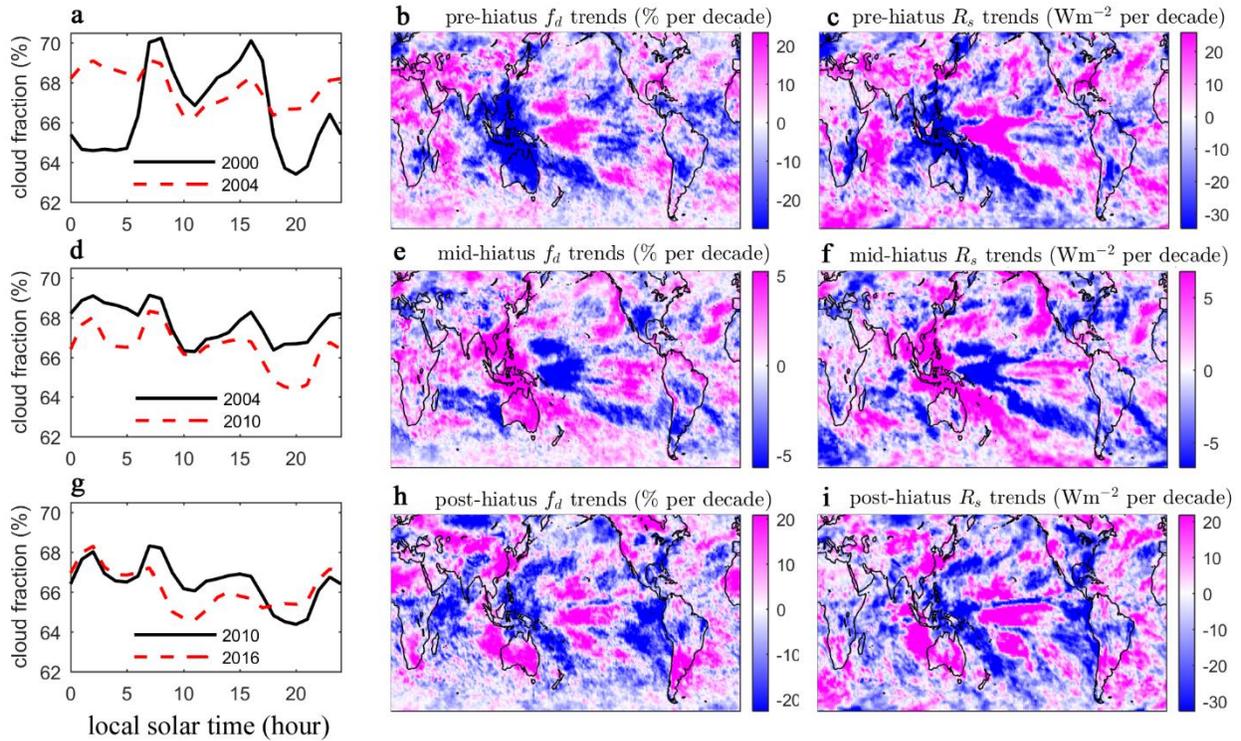

Fig. 2 Trends of daily cloud cycle, daytime cloud fraction, and shortwave radiative fluxes. The left column shows the DCC transition during (**a**) pre-hiatus (**d**) mid-hiatus, and (**g**) post-hiatus periods. The DCC is averaged over the entire year at local time of 1-hour interval for each grid point and then averaged over the globe. The middle column shows the trends of daytime cloud fraction during (**b**) pre-hiatus (**e**) mid-hiatus, and (**h**) post-hiatus periods. The right column shows the trends of outgoing TOA shortwave radiative fluxes during (**c**) pre-hiatus (**f**) mid-hiatus, and (**i**) post-hiatus periods. More trends of longwave radiative fluxes and nighttime cloud fraction are shown in Supplementary Fig. 3. The DCC was derived from CERES SYN; cloud fraction trends were derived from CERES SSF; radiative flux trends were calculated from CERES EBAF (see Methods).

The consistent DCC-radiation patterns found in Fig. 2 corroborate our hypothesis that DCC, in pace with the ocean circulation, has a strong impact on the global energy budget. However, evaluating such impacts using satellite observations is quite challenging: in fact, while their geographical patterns are strong and clear, globally they tend to cancel each other, resulting in very subtle trends in the global energy balance (see Supplementary Fig. 4)[17,18]. To reduce this uncertainty when using satellite observations, we calculated the relative change of clouds at any time of the day to analyze the cloud cycle

$$\tilde{f}(t) = \frac{f(t) - f_0}{f(t)}, \quad (1)$$

where $f_0$ is daily mean cloud fraction. Such a metric is similar to the diurnal asymmetrical ratio, which has been used for satellite calibration to provide best estimate of radiative fluxes (see Methods). Using relative ratios, rather than absolute values to describe the daily cycles, helps reduce the impact of artifacts from geostationary satellites[18]. The relative change of cloud fraction $\tilde{f}$ can then be readily linked to the radiative fluxes at the top of the atmosphere through the cloud radiative effects (CRE). It has been shown that the DCC radiative effects (DCCRE) in terms of cloud fraction can be defined as (see reference 26 and Methods)

$$\text{DCCRE} = \tilde{f}\text{CRE}. \quad (2)$$

The change of DCCRE quantifies the radiative impacts of the detailed DCC variations, including its phase shift and amplitude modulation[26].

When the global averages of DCCRE at each grid point (see Methods) are compared with the global mean surface temperature (see Fig. 3), the trends show striking similarity, with a slow and steady decrease during the mid-hiatus period and a fast increase during the pre- and post-hiatus periods. The pre-hiatus increase accelerates the global warming trends and mid-hiatus decrease counteracts the warming effects to maintain a relatively constant global temperature. Particularly interesting is the sudden increase in 2016, the hottest year on record, with an increase of about 0.5 Wm$^{-2}$ in net radiative flux by DCC variations (in this case, the estimates of radiative impacts are made more reliable thanks to the newest generation of geostationary satellites such as Himawari-8[18]). These trends are primarily accounted for by the fluctuation of $\tilde{f}$ as the recalculated DCCRE without year-to-year variation of CRE (see Methods) also show the similar trends (thin lines in Fig. 3).

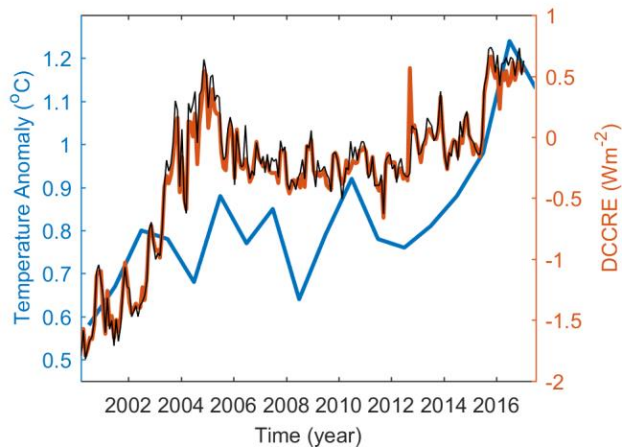

Fig. 3. Comparison of global mean surface temperature anomaly and daily cloud cycle radiative effects (DCCRE). The thin black lines are the recalculated DCCRE without year-to-year variations of CRE (see Methods). The temperature data (thick blue lines) are from the National Aeronautics and Space Administration Goddard Institute for Space Studies and the DCCREs (thick brown lines) are calculated from CERES SYN (see Methods).

The similarity in the DCCRE and temperature trends is strongly suggestive of the existence of a feedback loop between them, whereby the ocean circulation affects the patterns of sea surface temperature, which influence the lower-tropospheric stabilities and large-scale atmospheric circulation[27–29]. These changes in turn may influence cloud dynamics inducing changes in daytime and nighttime clouds[30–32], whose impacts on the Earth's energy balance end up reinforcing the patterns of sea surface temperature (see Fig. 2 and Supplementary Fig. 3). As a result, the climate system remains 'locked' for an extend period of time, in a hiatus-type of behavior, giving rise to a staircase-like function of global mean surface temperature[12].

In summary, a strong correlation of the daily cycle of cloud fraction with the pacific decadal oscillation may play a strong role in adjusting the Earth's energy budget and global mean surface temperature. This may have important implications for future climate predictions: while previous studies on climate hiatus have suggested that the imbalanced energy is stored in the deep oceans, the present results suggest that part of the imbalanced energy may actually be removed from the Earth's balance by the DCC amplitude modulations. Further investigations using other cloud properties, when they will become available, will provide more comprehensive assessment of the DCC radiative impacts on climate variability. Clarifying these climate processes at finer temporal resolutions could shed light into the difficult problem of disentangling the impacts of anthropogenic activity and nature variability on climate change.

**Methods**

**Daytime and Nighttime Cloud Properties**

The data product of Single Scanner Footprint (SSF) from Clouds and the Earth's Radiant Energy System (CERES)[18] provides global mean cloud fraction averaged over the entire day, $f_0$, and also over the daytime, $f_d$. These data can be used to calculate the nighttime cloud fraction, $f_n$, being

$$f_0 = \frac{f_d + f_n}{2}. \tag{3}$$

We also consider the daily amplitude of cloud fraction

$$f_a = \frac{f_n - f_d}{2}. \tag{4}$$

The global relationships between $f_d$, $f_n$, $f_a$, $f_0$ and PDO are shown in Fig. 1.

## Daily Cloud Cycle Radiative Effects (DCCRE)

The radiative flux at the top of the atmosphere $R$ depends on time of the day $t$, cloud fraction $f$, and any other related climate variables $x$. This flux $R(f,x,t)$ can be split into a contribution due to mean cloud fraction $f_0$ and the daily fluctuation ($f_{DCC} = f(t) - f_0$). The former can be expressed as $R(f_0, x, t)$, while the latter is the DCC radiative effects (DCCRE)[26]

$$\text{DCCRE} = R(f,x,t) - R(f_0,x,t). \tag{5}$$

The thick brown line in Fig. 3 shows the variations of global mean DCCRE, which quantifies the total radiative impacts of DCC phase shift and amplitude modulation[26]. The calculation of DCCRE from Eq. (5) requires $R(f,x,t)$ and $R(f_0,x,t)$, both of which are linked to cloud radiative effect (CRE) method, defined as[33–36]

$$\text{CRE} = R(f,x,t) - R_{clr}(x,t) = f(t)[R_{cld}(x,t) - R_{clr}(x,t)], \tag{6}$$

where $R$, $R_{cld}$ and $R_{clr}$ are all-sky, cloudy-sky, and clear-sky radiative fluxes at the top of the atmosphere. Solving (6) for $R$ gives

$$R(f,x,t) = f(t)R_{cld}(x,t) + [1-f(t)]R_{clr}(x,t); \tag{7}$$

that is, the all-sky $R$ is the sum of $R_{cld}$ and $R_{clr}$ weighted by total cloud fraction $f$. Taking $f = f_0$ into (7), one obtains

$$R(f_0,x,t) = f_0 R_{cld}(x,t) + (1-f_0)R_{clr}(x,t). \tag{8}$$

Combining (5), (6), (7), and (8) yields

$$\text{DCCRE} = \tilde{f}(t)\text{CRE}, \tag{9}$$

where $\tilde{f}(t) = [f(t) - f_0]/f(t)$ is the relative change of total cloud fraction (also see Eq. (1) in the text). It suggests both the relative variation of DCC ($\tilde{f}$) and CRE contribute to DCCRE. To investigate which term has a stronger contribution, we recalculated DCCRE without the year-to-year variations of CRE

$$\text{DCCRE}' = \tilde{f}\text{CRE}'. \tag{10}$$

where CRE' is the long-term mean daily cycle of CRE. Fig. 3 shows that the DCCRE' (thin black lines) are close to the DCCRE (thick brown lines), suggesting $\tilde{f}$ is the primary contributor to the DCCRE variations.

## Diurnal Correction

The diurnal correction[18] is a method used to abstract diurnal information from geostationary satellites without including their artifacts. To do so, a diurnal asymmetry ratio is defined as

$$r_a = \frac{R_{am} - R_{pm}}{R_0}, \quad (11)$$

where $R_{am}$, $R_{pm}$, and $R_0$ are the mean shortwave radiative fluxes before noon, after noon, and for the entire day, respectively. This metrics is used to link the data product of Synoptic fluxes and clouds (SYN) and Single Scanner Footprint (SSF) to provide data product of Energy Balanced and Filled (EBAF) with best estimation of radiative fluxes[18]. Such a metric is similar to the relative change of cloud fraction (see Eq. (1)); both of them use ratios, rather than absolute values, to describe the daily cycles, to reduce the impacts of artifacts from geostationary satellites[18].

## Data availability

Clouds and radiative fluxes data are from Clouds and the Earth's Radiant Energy System (CERES) Single Scanner Footprint (SSF), Synoptic fluxes and clouds (SYN), and Energy Balanced and Filled (EBAF) (https://ceres.larc.nasa.gov/). PDO index data were downloaded from NOAA National Centers for Environmental Information (https://www.ncdc.noaa.gov/teleconnections/pdo/). Earth's surface temperature data were obtained from NASA Goddard Institute for Space Studies (GISS) (https://data.giss.nasa.gov/gistemp/).

## Acknowledgments

We acknowledge support from the USDA Agricultural Research Service cooperative agreement 58-6408-3-027; and National Science Foundation (NSF) grants EAR-1331846, EAR-1316258, FESD EAR-1338694 and the Duke WISeNet Grant DGE-1068871. The useful suggestions offered by three anonymous reviewers are gratefully acknowledged.

## Author contributions

A.P. and J.Y. conceived and designed the study. J.Y. wrote an initial draft of the paper, to which both authors contributed edits throughout.

Supplementary Material for

# Reinforcement of Climate Hiatus by Decadal Modulation of Daily Cloud Cycle

Jun Yin[1,2], Amilcare Porporato[1,2*]

[1]Department of Civil and Environmental Engineering, Princeton University, Princeton, New Jersey, USA.

[2]Princeton Environmental Institute, Princeton University, Princeton, New Jersey, USA.
*Correspondence to: aporpora@princeton.edu

The following figures provide complementary information regarding the details of the analyses on the daily cycle of clouds and its linkage to the Earth's energy balance:

- Supplementary Fig. 1 compares the daily and seasonal cycles of cloud radiative effects.
- Supplementary Fig. 2 compares the time series of PDO index and global mean surface temperature.
- Supplementary Fig. 3 compares the trends of nighttime clouds and longwave radiative fluxes at the top of the atmosphere (a complementary information to Fig. 2 in the main text).
- Supplementary Fig. 4 shows the trends of radiative fluxes at the top of the atmosphere from different versions of CERES EBAF data.

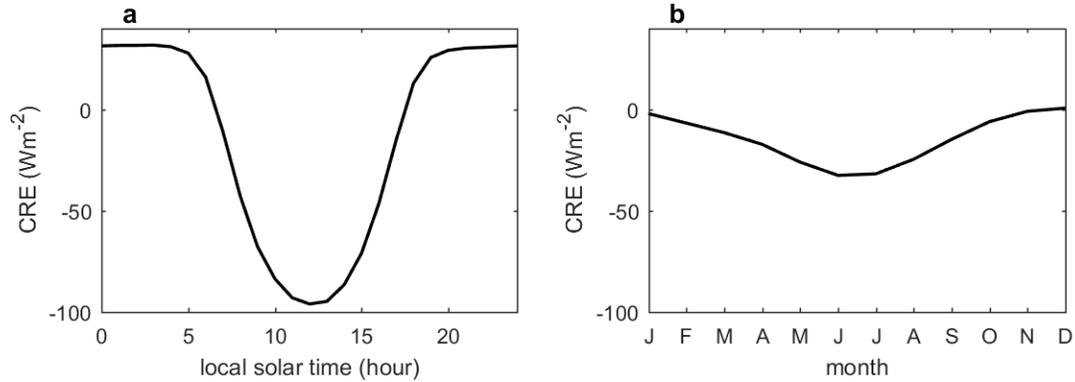

Supplementary Fig. 1. Comparison of daily and seasonal cycles of clouds. The (a) daily and (b) seasonal cycles of cloud radiative effects (CRE) averaged over the North Hemisphere during March 2002 – February 2017. Data are from CERES SYN (see Methods).

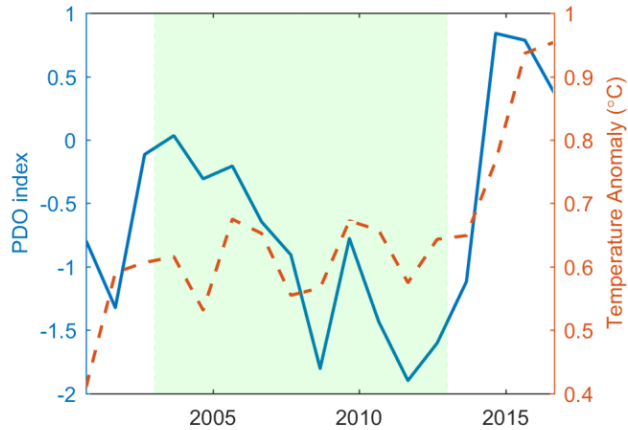

Supplementary Fig. 2. Comparison between global mean surface temperature anomaly and POD index. The temperature data are from NASA Goddard Institute for Space Studies (GISS) and PDO index are from NOAA National Centers for Environmental Information. The shaded area divides the early 21st century into pre/mid/post-hiatus periods.

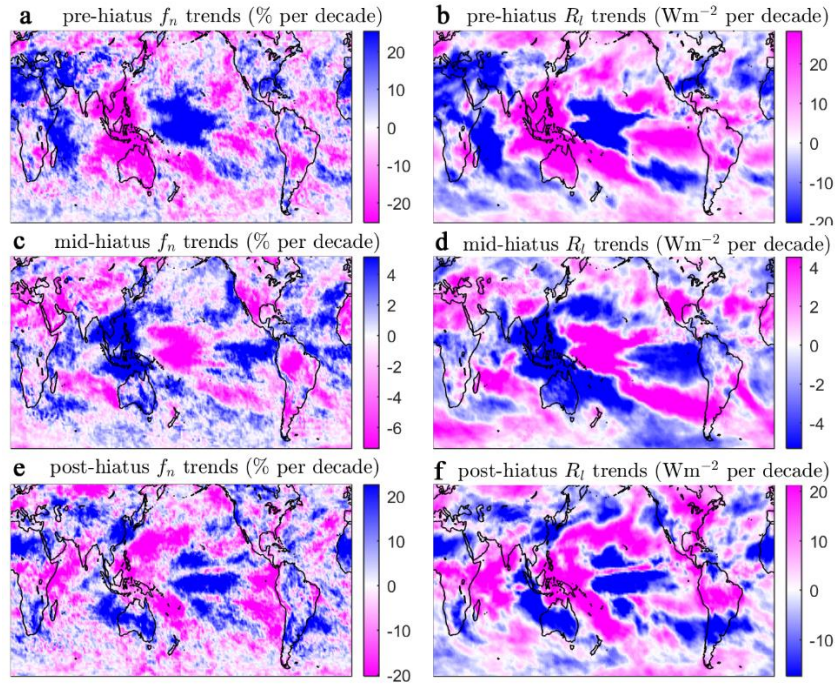

Supplementary Fig. 3. Trends of nighttime clouds and longwave radiative flux at the top of the atmosphere. The trends of (a, c, and e) nighttime cloud fraction, $f_n$, have a similar spatial pattern as the trends of (b, d, and f) longwave radiative fluxes at the top of the atmosphere, $R_l$. The trends at (a and b) pre-hiatus and (e and f) post-hiatus periods are usually opposite to the trends at (c and d) mid-hiatus period. Cloud fraction trends were derived from CERES SSF; radiative flux trends were calculated from CERES EBAF (see Methods).

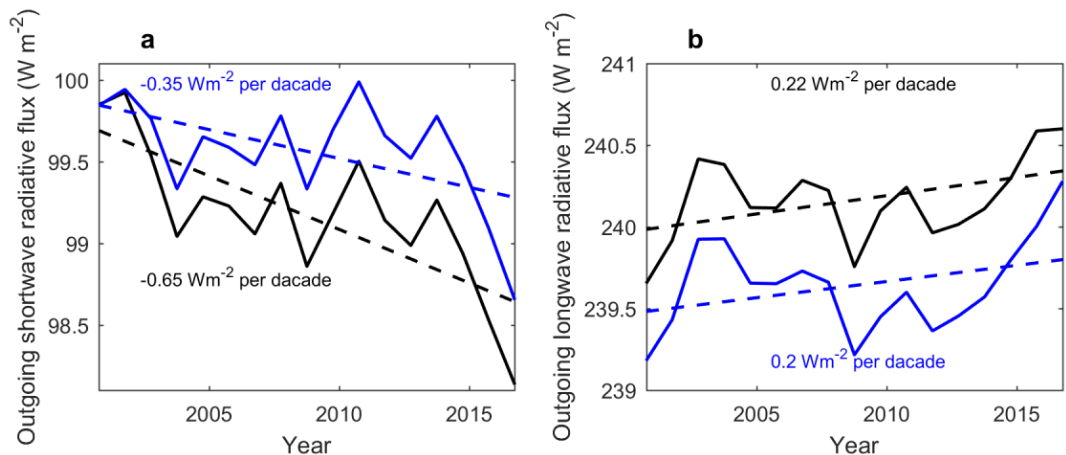

Supplementary Fig. 4. Global mean outgoing (a) shortwave and (b) longwave radiative fluxes at the top of the atmosphere. The blue and black lines are from CERES EBAF Ed 2.8 and Ed 4.0, respectively. The dash lines are the corresponding linear fit, showing the long-term trends.